# Integration of Porous Graphene and 3D-printed Piezopolymer for Flexible Ultrasound Transducers

**Shirin Movaghgharnezhad [1,\*], Ehsan Ansari [1], Clayton A Baker [1], Ahmed A Bashatah [1], Dulcce A Valenzuela [1], Pilgyu Kang [2], Parag V Chitnis [1,\*]**

[1]Department of Bioengineering, George Mason University, Fairfax, VA 22030, USA

[2]Department of Mechanical Engineering, George Mason University, Fairfax, VA 22030, USA

\* Corresponding authors.

*E-mail addresses:* Shirin Movaghgharnezhad (smovaghg@gmu.edu), Parag V Chitnis (pchitnis@gmu.edu)

**Ultrasound technology is crucial in diagnostic imaging, making it widely used in medical applications. However, traditional ultrasound transducers face limitations in flexibility and ease of fabrication, leading to the exploration of thin-film and flexible piezoelectric materials. Here, we present a novel approach that combines laser graphitization with 3D printing to integrate flexible laser-induced porous graphene (LIG) with poly(vinylidene fluoride-trifluoroethylene) (PVDF-TrFE), resulting in the development of flexible LIG/PVDF-TrFE ultrasound patches. The thickness of PVDF-TrFE is adjusted to tune the central frequency of the ultrasound transducer, allowing customization within a range of 10 to 28 MHz. LIG-based ultrasound transducer demonstrates a high signal amplitude of 6.72 V and a signal-to-noise ratio (SNR) of 433, along with a −6 dB bandwidth of 8.86 MHz (37%). The LIG-based transducer exhibits higher acoustic performance compared to the smooth silver-based transducer. A high-quality two-dimensional ultrasound image, including a B-mode image of a cyst phantom, demonstrates the transducer's imaging capabilities. The patterning of LIG-based electrodes facilitates the desired sensor configuration, demonstrating the suitability of our novel technique for producing flexible transducer arrays without dicing and cutting. The materials cost of our LIG/PVDF-TrFE transducer is under $5 per unit, making it a low-cost solution for ultrasound patches.**

# 1. Introduction

Diagnostic ultrasound imaging is a versatile and indispensable tool in clinical care, including anatomical and fetal imaging, echocardiography and Doppler ultrasound for assessing vascular health, and musculoskeletal ultrasound for assessing orthopedic conditions and injuries (**Fig. 1a**). Ultrasound imaging relies on transmitting ultrasound pulses into the medium and detecting reflected echoes from different tissue interfaces. Therefore, ultrasound imaging requires an ultrasound transducer (UST) that can efficiently convert electrical energy into mechanical (acoustic pressure) energy, and vice versa (i.e., piezoelectric material) [1] Until recently USTs have been fabricated using lead zirconate-titanate (PZT) ceramics [2], lead perovskite single crystals [3], and PZT composites. Despite their strong piezoelectric performance, these materials present several inherent challenges: their acoustic impedance differs significantly from that of tissue, leading to poor energy transfer and reduced imaging quality, necessitating the use of intermediate matching layers for efficient coupling; their rigidity and brittleness make them difficult to machine, complicating the fabrication of small elements in array transducers; this also makes them an unappealing choice for conformable devices [4], The increasing demand for flexible and wearable ultrasound devices has motivated researchers to explore innovative thin film and flexible materials, and techniques to miniaturize and enhance the performance of transducers.

Recent studies have demonstrated tremendous potential of wearable USTs in biomedical applications that require chronic imaging and sensing [5], [6], [7], [8], [9], [10], [11], however, these prior works still used piezoceramics that suffer from challenges associated with rigid and brittle materials. Piezopolymers, which are amenable to fabrication into thin and flexible films, are complementary alternative piezoceramics for producing USTs with enhanced flexibility, while providing an added advantage of better acoustic-impedance matching with tissue. For example, ~~Du et al. reported polymer-based optoacoustic transducers for ultrasound imaging using a composite of lead halide perovskites and polydimethylsiloxane (PDMS) [12].~~ Gijsenbergh et al. demonstrated polymer-based piezoelectric micromachined ultrasound transducers (PMUT) using polyvinylidene fluoride-trifluoroethylene (PVDF-TrFE) for simple gesture recognition [12]. PVDF is a piezoelectric polymer with properties amenable to fabricating flexible USTs, such as a wide frequency bandwidth for fine-resolution imaging, mechanical flexibility for wearable and conformable devices, acoustic impedance quite close to that of tissue for improved acoustic coupling with tissue, biocompatibility and lightweight for facilitating chronic applications, low cost, and ease of fabrication [13]. The copolymer of PVDF-TrFE expands on these beneficial characteristics by providing high polarization, excellent compatibility with solvents, a high electromechanical coefficient, superior thermal stability, and notably minimal mechanical and electrical loss when compared to standard PVDF [14], [15]. In another study, the integration of ultrafine β-nickel hydroxide (β-Ni(OH)$_2$) nanoplates with PVDF-TrFE improved the polarization and piezoelectric response, due to the increased effective surface area of piezopolymer, producing a high-frequency UST (50 MHz) with 100% bandwidth, which is very useful for fine-resolution imaging of living tissues [16]. The high flexibility, high piezoelectric properties, and easy polarization of PVDF-TrFE make it a potential option to be used in designing flexible USTs that are potentially disposable and highly amenable for hands-free applications such as assessing dynamic muscle function during physical activity and exercise, chronic monitoring of blood flow, etc.

Fabrication of flexible USTs necessitates the integration of piezoelectric material with conductive electrodes on a flexible substrate. Conventional piezoceramic materials often require high-temperature processes (over 700 ºC) or methods like mechanical dicing, which are not amenable for integration with flexible substrates [17]. Although strategies have been developed to address this challenge, they represent a complex fabrication process. One approach involved dice-and-fill technique for fabricating 1-3 piezoelectric composites on rigid glass substrates, which were then transferred to flexible Ecoflex substrates [5]. Another work fabricated flexible USTs by depositing Pt/PZT/Pt structured layers on silicon wafers and subsequently transferring them to polyimide films [18]. Other research reported flexible piezoelectric transducers based on AlN thin films, with direct deposition of a Mo/AlN/Mo structure on a flexible polyimide (PI) substrate through the sputtering method [19]. However, these methods produce devices with stacked layers, which are prone to delamination and low mechanical robustness. Therefore, achieving reliable flexible USTs requires innovative advancements in piezoelectric materials and their seamless integration with flexible conductive electrodes.

Graphene is a promising material for fabricating functional UST electrodes as it exhibits excellent characteristics, including mechanical robustness [20], flexibility [21], high charge carrier concentration and mobility [22], thermal conductivity [23], and chemical resilience [24]. With these outstanding properties, the utilization of graphene-based nanocomposites has enabled creation of novel electronic devices and sensors. For instance, single-layer graphene (SLG) synthesized through chemical vapor deposition (CVD) on Si/SiO2 calibration grating substrates was found to exhibit a significant piezoelectric effect, with a d$_{33}$ piezoelectric coefficient of 1.4 nm V$^{-1}$, for fabrication of energy-harvesting devices [25]. Other studies showed that PVDF/graphene-based nanocomposites [26], [27], [28], [29], [30] offer enhanced piezoelectric performance due to the synergistic effect of improved electrical properties and enhanced interface interaction between the graphene and PVDF owing to the advancement in the synthesis process, such as better dispersion of graphene within the PVDF matrix and improved bonding at the interface [30][32]. However, the fabrication process of graphene substrate in these studies is complicated and time-consuming. Therefore, the integration of piezopolymers with graphene for fabricating ultrasound-imaging transducers remains unexplored, primarily due to a lack of a facile and scalable method for producing graphene-based conductive substrates that can facilitate seamless integration with piezopolymers.

In this study, we present an innovative approach for fabricating flexible ultrasound patches based on laser-induced graphene (LIG) and additive manufacturing of piezopolymer via 3D printing of PVDF-TrFE ink. LIG provides excellent electrical conductivity, mechanical flexibility, and ease of patterning, making it ideal for creating durable and conformable electrodes. PVDF-TrFE films are flexible and have an acoustic impedance that is closely matched to that of tissue. The LIG is synthesized from a polyimide (PI) precursor through a photothermal laser process [31], [32], [33], [34], [35], [36] that uses a $CO_2$ laser to convert PI into porous graphene in a single step. PVDF-TrFE ink can be printed directly onto the graphene pattern, where a portion of it infiltrates the interconnected pores of the LIG, forming a strong interfacial bond (**Fig. 1b**). This infiltration enhances adhesion between the piezopolymer and the electrode, improving mechanical stability. The integration with LIG's porous structure increases interaction at the LIG/PVDF-TrFE interface, expanding the effective surface area of the LIG electrode and enhancing the acoustic performance of the UST. **Fig. 1c** illustrates a schematic representation of the key components of the LIG-based UST and **Fig. 1d** shows the UST on a forearm. The PVDF-TrFE thickness could be tuned to achieve a target center frequency, with our devices covering a range from 10 MHz to 28 MHz. This tunability allows for adaptability in various medical applications, depending on the imaging or diagnostic requirements. Our technique for developing flexible LIG/PVDF-TrFE USTs is scalable, fast, and versatile in creating different patterns (**Fig. 1a**), with a manufacturing cost of less than $5 per device, rendering them disposable and economically viable for producing transducer arrays for B-mode imaging in medical applications.

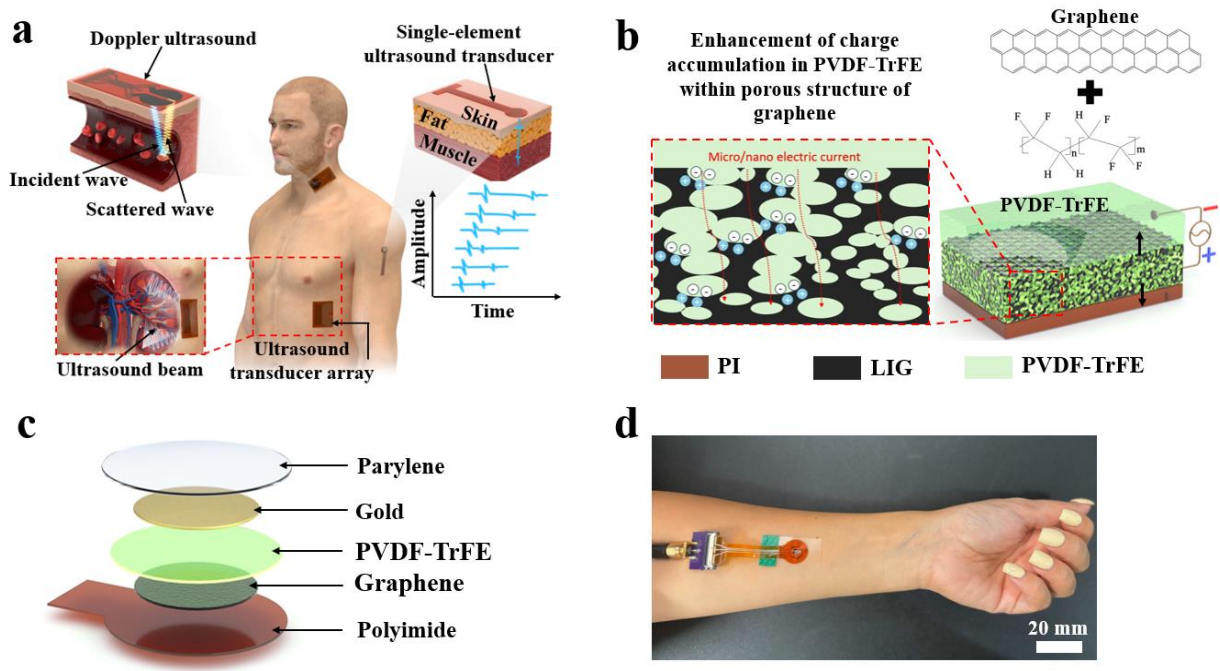

**FIGURE 1.** A flexible ultrasound transducer based on LIG/PVDF-TrFE composite. (a) Ultrasound has a wide range of applications, including assessing blood flow in major arteries using Doppler transducers, detecting muscle movement with single-element transducers, and evaluating kidney health with transducer arrays. (b) Schematic illustrating the charge enhancement mechanism and improved piezoelectricity resulting from the integration of a piezopolymer with porous graphene. The porous graphene structure increases the surface interaction area with PVDF-TrFE as the polymer embeds within it, facilitating charge accumulation and enhancing piezoelectric performance. (c) Schematic illustration of the layered design, highlighting the main component of the LIG/PVDF-TrFE transducer. (d) Photograph of the single-element LIG/PVDF-TrFE transducer placed on a forearm.

**Results and discussion**

*Fabrication of flexible LIG/PVDF-TrFE transducers*

To synthesize composite LIG/PVDF-TrFE, a combination of photothermal laser manufacturing and printing techniques was employed (**Fig. 2a**). Initially, a single LIG element, 3.5 mm in diameter, from commercial polyimide film (Dupont Kapton HN, 5Mil thick) was synthesized using photothermal laser treatment. This single element becomes the functional working electrode (Materials and methods). The use of photothermal laser processing presents a quick and simple approach in contrast to conventional methods such as chemical vapor deposition (CVD), enabling versatile patterning and targeted graphitization. The light intensity used in this process is specifically adjusted for the material undergoing laser treatment, controlling the local temperature through various operational parameters including laser power, laser speed, and pulse per inch (PPI). This technique involves the use of a 10.6 μm $CO_2$ laser, where the area of focus reaches temperatures exceeding 2500 ºC. This extreme heat breaks down the covalent bonds in PI, leading to the release of $H_2$, $N_2$, and $O_2$ gases [31], which create a variety of macro- and mesopores within the graphene's structure [32]. The precise control of laser settings enables fine-tuning of graphitization depth, pore size, and conductivity for device optimization. We used optimal laser settings of 4.2 W power, 3.5 in s$^{-1}$ speed, and PPI 1000 (Materials and methods).

Subsequently, conductive silver ink was used to print interconnects and contact pads using a nozzle-based dispensing printer. This was followed by 3d-printing of piezoelectric PVDF-TrFE to achieve the desired thickness (ranging from 50 to 100 μm) of piezopolymer deposition (Materials and methods). The printed PVDF-TrFE layer was dried and annealed at 140°C. After the annealing process, the PVDF-TrFE was subjected to a corona poling process at a high DC voltage for 100 minutes (further details can be found in Materials and methods). This step aligns the dipole molecules within the material, enhancing its piezoelectric properties [15]. Finally, a 60-nm gold (Au) layer was deposited by ion beam sputtering to serve as a top electrode for grounding and a dielectric parylene C layer was applied for encapsulation. With a total thickness of ~170 μm, the transducer is inherently bendable due to the high flexibility of constituent layers (**Fig. 2b**). This flexibility is crucial for wearable acoustic imaging applications, as it allows the transducer to conform to irregular surfaces, improving contact and image quality in dynamic, real-world conditions. Notably, the scalability and precise patterning capability of the photothermal laser manufacturing technology enable the fabrication of USTs with different geometries and configurations, which included an M-shaped UST 7-mm across, a dual-element doppler transducer (**Fig. 2c**), and a 32-element array with ~390-μm pitch (Supplementary **Fig. S1a**). We measured the resistance of the elements after cyclic bending ($\varepsilon_b$ = 4.5%) and twisting (from − 90° to + 90°) for up to 10,000 cycles, confirming their mechanical flexibility and durability (**Fig. S1b**). The resistance and capacitance of all 32 elements were also measured, and the variation between elements was negligible (**Fig. S1c**).

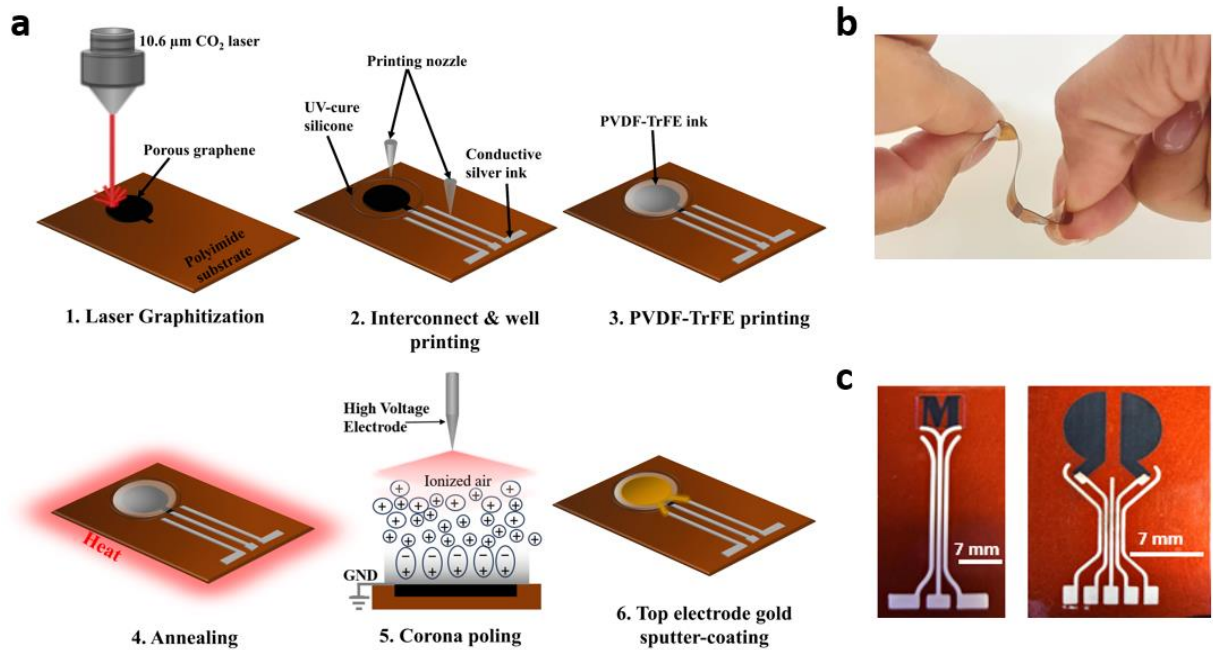

**FIGURE 2.** LIG/PVDF-TrFE ultrasound transducer. (a) Schematics of the fabrication process of LIG/PVDF-TrFE composite. (b) Fabricated ultrasound transducer illustrating its mechanical flexibility. (c) LIG-based transducers with different patterns, prior to PVDF-TrFE deposition.

*Materials characterization of LIG and LIG/PVDF-TrFE composite*
The morphological and crystalline structure of the LIG and LIG/PVDF-TrFE composite was analyzed using scanning electron microscopy (SEM), Raman spectroscopy, and X-Ray diffraction crystallography (XRD). The SEM (**Fig. 3a**) reveals the porous structure of LIG with an average diameter in the sub-micrometer range, and a 50 μm thick LIG is displayed in the cross-sectional SEM image (**Fig. S3**). This porous structure enables PVDF to embed within the LIG, strengthening the bond at the interface and reducing delamination. Additionally, the increased effective surface area of PVDF enhances piezoelectricity, further improving the transducer's performance in wearable applications. The PVDF-TrFE layer on LIG displays a rough surface with a fibrous pattern, attributed to the fabrication method solution-based nozzle printing (**Fig. 3b**). The cross-sectional SEM image reveals the PVDF-TrFE embedded within LIG as well as coated uniformly on top (**Fig. 3c**). The average thickness of the PVDF-TrFE is adjusted via the mass of the solution deposited, increasing from 50 μm (**Fig. 3c (i)**) at 15 mg deposition to 100 μm-thick (**Fig. 3c (iv)**) at 30 mg. These results demonstrate not only the effective integration of LIG with PVDF-TrFE but also the capability to tune the piezoelectric thickness for the desired resonance frequency of the ultrasound transducer. This tuning allows for improved control over image resolution, with higher frequencies providing finer detail, making the transducer adaptable for different medical imaging needs.

The Raman of LIG exhibits three prominent peaks: D (≈1346 cm$^{-1}$), G (≈1576 cm$^{-1}$), and 2D (≈2688 cm$^{-1}$) (**Fig. 3d**), indicating the formation of graphene through the photothermal laser process, which is consistent with previous literature [31], [32]. The high $I_G/I_D$ ratio of 1.83, along with a sharp G peak, signifies pronounced graphitization. The large crystalline size ($L_a$) of 30 nm along the axis of graphitic structures, as inferred from the $I_G/I_D$ ratio, verifies both the high degree of graphitization and superior crystallinity. XRD pattern of the LIG exhibits a characteristic (002) peak at 26.5° corresponding to a d-spacing of ~0.34 nm. The peak at 44° associated with the (100) graphitic crystal phase demonstrates the presence of graphitic structure on the LIG [31]; the broadness of this peak is attributed to the diffraction from the PI substrate (**Fig. S2**). The crystalline properties of PVDF-TrFE after annealing and poling were characterized by XRD. **Fig. 3e** shows a sharp peak at 20.12° represents the Bragg diffraction of (110)/(200) of the β-phase [37], which is indicative of the ferroelectric β-phase with an all-trans conformation. After poling, the peak at 20.12° became sharper, indicating an increase in piezoelectric properties. Similar characterization was performed on poled PVDF-TrFE atop a smooth silver electrode (Ag/PVDF-TrFE) to compare its piezoelectric properties with LIG/PVDF-TrFE and investigate the effect of piezopolymer integration with porous graphene (**Fig. 3f**). The sharper β-phase peak of LIG/PVDF-TrFE indicates enhanced crystallinity and improved polarization, which are structural features known to support better piezoelectric behavior. This improvement is attributed to the increased effective surface area of PVDF-TrFE in the LIG/PVDF-TrFE composite, resulting from the porous nature of LIG. In this structure, the porous LIG acts not only as a conductive electrode but also provides a textured interface into which the lower region of the PVDF-TrFE layer embeds. This increased interfacial contact area promotes stronger mechanical anchoring and improved coupling between the piezopolymer dipoles and the electrode surface, facilitating more efficient charge transfer. This expanded interface enhances charge collection and distribution, leading to improved energy conversion and stronger electrical signals. In contrast, the smooth surface of Ag limits this interaction, reducing piezoelectric performance. Piezoelectric performance, in this context, refers to the material's ability to convert mechanical stress into electrical energy, with key indicators including the piezoelectric coefficient and electromechanical coupling efficiency. Cross-sectional SEM images of LIG/PVDF-TrFE and Ag/PVDF-TrFE confirm the infiltration of PVDF into the porous structure of LIG, in contrast to its formation solely on top of silver, which increases the risk of delamination (**Fig. S4**). By leveraging LIG's porosity to increase its interaction with PVDF-TrFE, we achieve more efficient energy conversion and stronger pulse-echo signals. This material engineering approach improves ultrasound transducer performance by enabling devices that are more sensitive, flexible, and lightweight. These advancements are particularly beneficial for high-resolution imaging, wearable technologies, and real-time, non-invasive diagnostics.

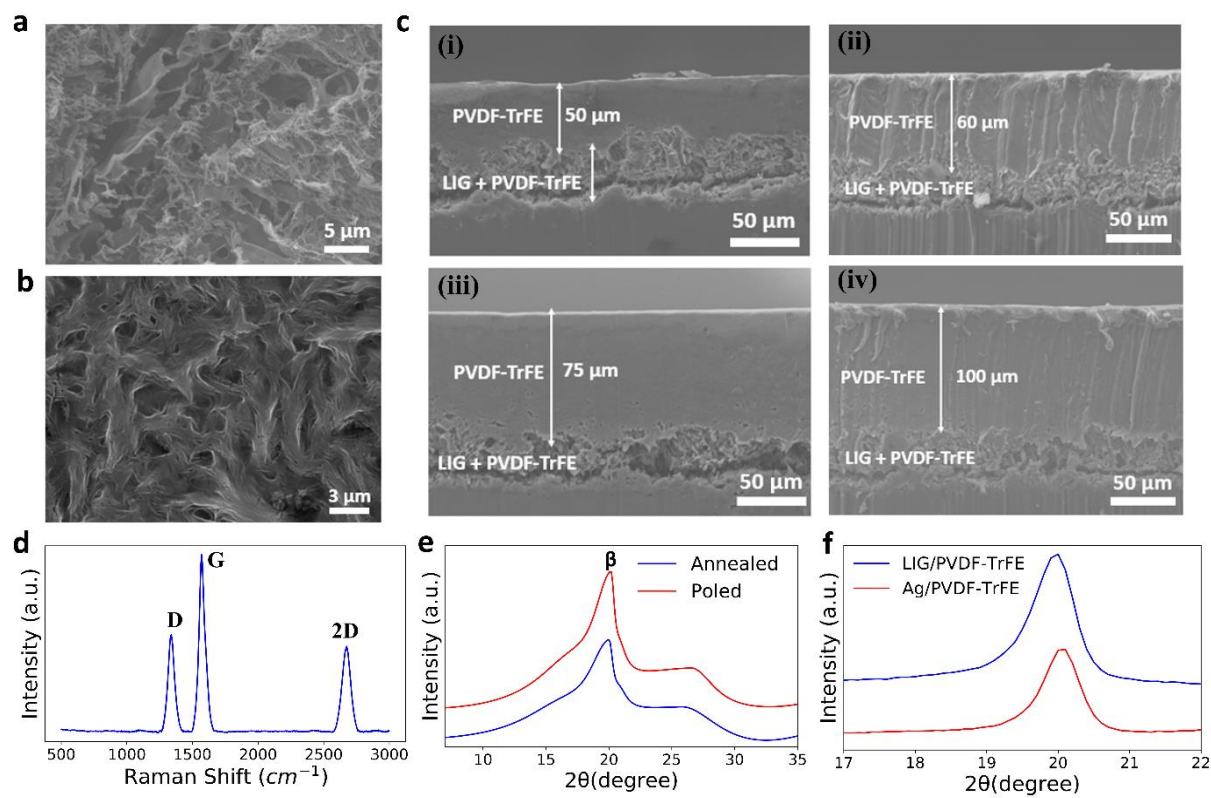

**FIGURE 3.** Morphological characterization of LIG, PVDF-TrFE, and LIG/PVDF-TrFE composite. Structural morphology is captured in surface SEM images of (a) bare LIG and (b) LIG/PVDF-TrFE. (c) Cross-sectional SEM images of the LIG/PVDF-TrFE with different PVDF-TrFE thicknesses. (d) Raman spectra of LIG, fabricated at a laser power of 4.2 W, scan rate of 3.5 in s$^{-1}$, and PPI of 1000. (e) X-ray diffraction of annealed and poled PVDF-TrFE indicates the formation of β crystalline phase. (d) X-ray diffraction of PVDF-TrFE deposited on LIG and Ag. A sharper peak of LIG/PVDF-TrFE at 2$\theta$=20º (β) than Ag/PVDF-TrFE indicates its higher piezoelectricity.

*Piezoelectric, dielectric properties, and impedance analysis of LIG/PVDF-TrFE composite*
The electrical impedance, dielectric, and piezoelectric properties of the LIG/PVDF-TrFE composite were investigated to obtain its effective electromechanical coupling coefficient ($K_{eff}$) and piezoelectric coefficient ($d_{33}$). Permittivity is an inherent characteristic of a material, independent of its thickness. Structural changes, as previously demonstrated with XRD, are related to variations in the dielectric permittivity of the transducer before and after poling (**Fig. 4a**). The relative permittivity of the poled LIG/PVDF-TrFE composite is higher compared to a composite that has only been dried and annealed. This indicates that the poling process successfully aligns the dipoles, leading to a higher dielectric constant and enhancing the material's dielectric properties. **Fig. 4b** shows the transducer's electrical impedance and phase as a function of signal frequency. The resonance and antiresonance frequencies are 18 and 22 MHz respectively, resulting in a $K_{eff}$ of 0.30, which is similar to the value expected from the literature [14]. The resonance frequency directly impacts the efficiency of energy transfer and device sensitivity, while the separation between resonance and antiresonance suggests low energy loss, enhancing overall performance. A compression piezoelectric response test was conducted to measure the output voltage of the LIG/PVDF-TrFE under sinusoidal compression. **Fig. 4c** displays the voltage response of both unpoled and poled transducers with the poled transducer showing an amplitude of 20 mV, confirming successful poling and piezoelectric response. The voltage response of the LIG/PVDF-TrFE was compared with that of the Ag/PVDF-TrFE, with the LIG/PVDF-TrFE producing twice the voltage, demonstrating its superior piezoelectric response (**Fig. S5**). This higher amplitude, attributed to the enhanced piezoelectric response of LIG/PVDF-TrFE compared to Ag/PVDF-TrFE, further corroborates the previously demonstrated XRD results. The piezoelectric coefficient ($d_{33}$) of LIG/PVDF-TrFE transducers was electrically measured using a Berlincourt setup. **Fig. 4d** shows the effect of the applied poling electric field on the $d_{33}$ coefficient of the transducer. The electric field was varied as 140, 170, 200, 260, and 290 V $\mu m^{-1}$. As illustrated in **Fig. 4d**, the LIG/PVDF-TrFE transducer (50 μm-thick PVDF) exhibited a higher piezoelectric coefficient at an applied electric field of 170, reaching a maximum $d_{33}$ of 23 PC $N^{-1}$, which is consistent with that of commercial PVDF-TrFE films. A poling electric field exceeding 260 V $\mu m^{-1}$ tended to cause intermittent arc discharge, leading to damage in the PVDF-TrFE layer. While a higher electric field can generate more charged ions that accumulate on the piezopolymer surfaces, an excessively strong electric field increases the risk of coating damage and dielectric breakdown.

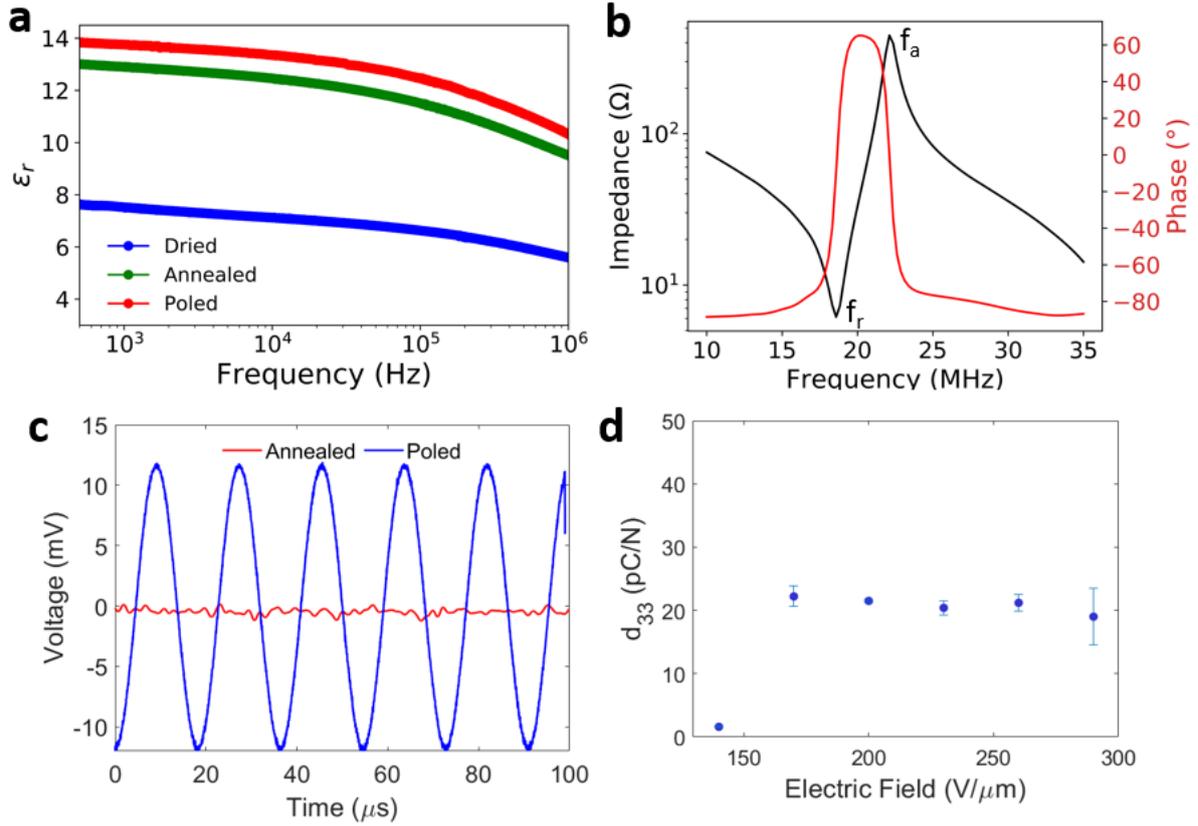

**FIGURE 4.** Dielectric and piezoelectric properties of the LIG/PVDF-TrFE composite. (a) Permittivity of un-poled and poled LIG/PVDF-TrFE. (b) Electrical impedance and phase of a LIG/PVDF-TrFE device. (c) Voltage response of the LIG/PVDF-TrFE under cyclic compression. (d) Piezoelectric coefficient of the poled LIG/PVDF-TrFE at different poling electric fields.

*Characterization of the acoustic output of LIG/PVDF-TrFE UST*

The acoustic tests were conducted to assess the bidirectional ultrasound-signal transduction capability of the LIG/PVDF-TrFE UST, which is critical for evaluating its performance in both transmitting and receiving ultrasound pulses. Graphene-based USTs were interfaced with pulse-echo electronics using a custom-designed printed circuit board for all acoustic tests (**Fig. S6**). The pulse-echo test was performed in water, using a setup displayed in **Fig. S7**, to evaluate the output acoustic performance of the LIG/PVDF-TrFE transducer (further details can be found in Materials and methods). **Fig. 5a** shows the pulse-echo signal captured by the LIG-based transducer, composed of a 50 µm-thick PVDF layer and a 125 µm-thick PI substrate. LIG-based transducer delivers a substantial peak-to-peak output voltage of 4.88 V, with a high SNR of 86.63. As the acoustic signal is generated symmetrically by the piezoelectric element, two pulses were observed from the impulse input: one due to the forward-propagated pulse (marked (1)), and the other one (marked (2)) due to the back-propagated pulse after it reflects off the near surface of the PI substrate (**Fig. 5a**). Based on the speed of sound in PI substrate (2142.3 m s$^{-1}$) and its thickness (125 μm), the delay between the pulses is calculated to be 0.12 μs, which is consistent with the time-domain signal (**Fig. 5a**). The output time-domain voltage signal alongside its frequency spectrum was analyzed. Using the Fast Fourier Transform (FFT), the time-domain signal was converted to its frequency domain, yielding a plot of relative amplitude (in decibels) against frequency (in MHz), as shown in **Fig. 5b**. The transducer exhibited a −6 dB (full-width half-max) bandwidth of 8.86 MHz (37%) with a center frequency of approximately 23.94 MHz. Although PVDF-TrFE is known to provide broadband response due to its favorable acoustic impedance and damping characteristics, the observed bandwidth is narrower than typically expected for polymer-based transducers. This is likely due to a combination of factors, including the use of a relatively thick PI substrate, which introduces internal reflections and multimode interference, and the porous LIG layer, which may lead to acoustic scattering and impedance mismatches at the LIG–PVDF-TrFE interface. Furthermore, the absence of a dedicated matching layer and damping backing structure contributes to a narrower −6 dB bandwidth. These factors will be considered in future device design to improve bandwidth while preserving the advantages of the LIG/PVDF-TrFE integration. We also fully acknowledge that LIG layer influences the overall acoustic impedance, resonance behavior, and vibrational coupling. Analytical modeling, such as using the Mason or KLM framework, would provide a valuable understanding of the multilayer interaction, especially in quantifying the acoustic impact of the LIG. Notably, the thickness of the PI substrate influences the resonance frequency, leading to the generation of odd harmonics [38], [39]. In particular, fabricating a 100 μm-thick PVDF layer with a 125 μm-thick PI substrate to achieve a ~10 MHz ultrasound transducer, based on the measured speed of sound in PVDF-TrFE (~2181 m/s), results in multiple resonance frequencies (**Fig. S8**). This occurs because the 125 μm-thick PI is insufficient to separate the forward and back-propagating pulses, leading to frequency-dependent interference between the two. Such interference can obscure the resonance peaks, making it challenging to achieve a clear and dominant center frequency. To mitigate this, we utilized a 500 μm-thick PI substrate, which significantly improved pulse separation (**Fig. S9**), resulting in a dominant first resonance peak at 10.05 MHz (**Fig. 5c**). By adjusting the PVDF thickness, we successfully tuned the central frequency down to 10 MHz (**Fig. 5d,** and **e**), demonstrating the potential of the LIG/PVDF-TrFE US transducer for operation within the medical ultrasound range (**Fig. S10**). LIG-based transducers exhibited reproducible output performance with an average voltage amplitude of 3.78, central frequency of 22.13 MHz, and bandwidth of 8.48 MHz (**Fig. S11**).

To assess the impact of integrating PVDF-TrFE with 3D porous micro-patterned LIG, we fabricated a control transducer by layering PVDF-TrFE onto a conventional smooth silver (Ag) electrode with dimensions matching those of the LIG-based device (3.5 mm aperture, 50 μm-thick PVDF-TrFE). This allowed a direct comparison of acoustic performance under structurally equivalent conditions. The flat surface of the Ag electrode offers limited interfacial contact with the piezopolymer, whereas the porous microstructure of the LIG electrode enables partial infiltration of PVDF-TrFE, increasing the effective electrode–polymer interface area. This structural difference enhances mechanical anchoring and local electric field concentration during poling, which promotes stronger dipole alignment, improved remnant polarization, and higher charge collection efficiency. Both devices were poled under the same electric field (170 V μm$^{-1}$), and their acoustic outputs were measured under identical pulse-echo settings (Materials and methods). The LIG/PVDF-TrFE transducer produced a signal amplitude more than three times that of the Ag/PVDF-TrFE device (**Fig. 5f**). This significant increase in output voltage reflects an enhancement in piezoelectric performance at the device level, arising from improved interfacial coupling and more effective polarization. The porous LIG structure enables stronger mechanical-electrical conversion by supporting better stress transfer and charge collection, which are critical for efficient piezoelectric signal generation. Further increases in signal amplitude and SNR, to 6.72 V and 433.46, respectively, were achieved by increasing the poling field to 260 V μm$^{-1}$ (**Fig. 5g**). A broader comparison with output voltage values reported for other ultrasound transducers is included in Table S1 for general reference (Table S1).

The transmit efficiency and received sensitivity of the LIG/PVDF-TrFE were measured by conducting hydrophone scan measurements. The experimental setup is displayed in **Fig. S12.** The device under test had a 3.5 mm circular aperture, composed of a 50 μm-thick PVDF-TrFE layer on a 125 μm-thick PI substrate, and further details are provided in Section 3.5. The transducer was driven with a 20 Vpp sine wave gated by a 0.5 μs rectangular window, with the frequency swept from 10 to 30 MHz in 250 kHz steps. Measurements were performed in water at a depth of 50 mm to ensure far-field conditions. The transmit transfer function was calculated by measuring the acoustic pressure generated at each frequency, while the receive sensitivity was derived from the transmit response using the reciprocity principle and electrical impedance of the device [40]. **Fig. 5i** shows both the transmit and receive transfer functions. The peak of the transmit transfer function was 2.05 kPa V$^{-1}$ at a resonance frequency of 19.85 MHz, and the peak of the receive transfer function was 0.044 V kPa$^{-1}$. Based on these values, the maximum pulse-echo efficiency was calculated to be 9%. The acoustic results of the LIG/PVDF-TrFE transducer indicate that its resonance tunability, combined with superior output voltage and high SNR, positions it as a promising candidate for medical ultrasound applications.

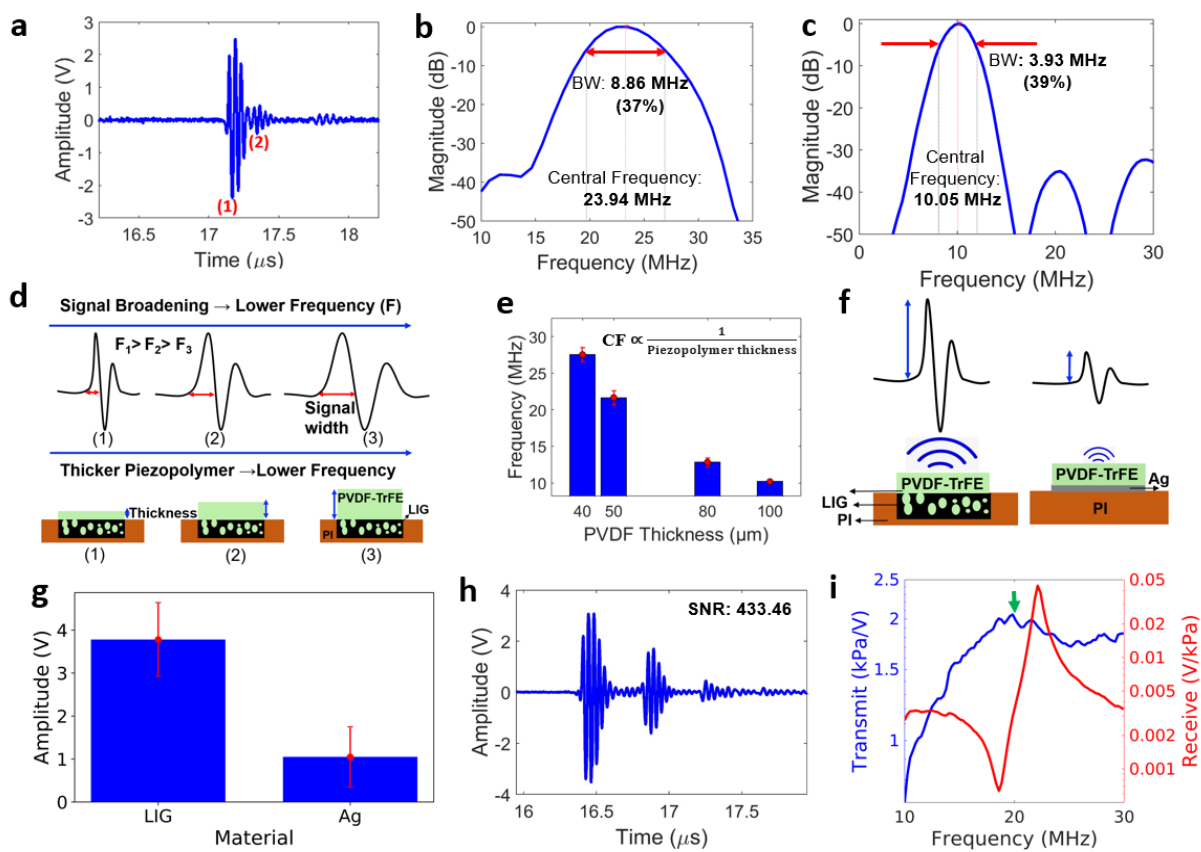

**FIGURE 5.** Acoustic performance of LIG/PVDF-TrFE ultrasound transducer. (a) Pulse-echo signal of the ultrasound transducer (forward-propagated pulse (marked (1), and back-propagated pulse (marked (2)). (b) The calculated frequency response of a LIG/PVDF-TrFE transducer with 50 μm PVDF-TrFE thickness and 125 μm-thick PI substrate. (c) The calculated frequency response of a LIG/PVDF-TrFE transducer with 100 μm PVDF-TrFE thickness and 500 μm-thick PI substrate. (d) Schematic illustrating the relationship between increasing PVDF-TrFE thickness and the reduction in central frequency, highlighting the corresponding signal broadening and transducer structure changes. (e) Resonance frequency of the LIG/PVDF-TrFE transducer at different PVDF-TrFE thicknesses. (f) Schematic illustrating the larger acoustic response of the LIG-based transducer compared to the Ag-based transducer. (g) Comparison of the acoustic output of the LIG-based and Ag-based transducers from pulser/receiver. (h) Pulse-echo signal LIG/PVDF-TrFE transducer. The transducer was poled at 260 V μm$^{-1}$. (i) Measured transmit and receive functions of LIG/PVDF-TrFE transducer.

To evaluate the imaging capabilities of the LIG/PVDF-TrFE UST, a series of imaging experiments were conducted using various targets in a controlled water tank environment. The first target was a 3D-printed reflector with the letter "G" extruded 5 mm from a rectangular block (**Fig. 6a**). The resulting image obtained from the normalized amplitude of the radiofrequency signals is depicted in **Fig. 6b**. A secondary image was produced (**Fig. S13**), highlighting the depth difference between the base and top layers of the letter "G,". This experiment successfully demonstrated the UST's ability to capture complex 3D shapes with high fidelity, highlighting its potential for detailed structural imaging. **Fig. S14** shows an ultrasound image of a 8 mm wide aluminum sheet embedded in a gel pad. The image clearly identifies the gel pad-air interface and multiple reflections from the aluminum sheet, demonstrating the transducer's capability to detect distinct reflective interfaces within a medium.

The second target, consisting of an 80 μm diameter wire affixed to a 3D-printed holder, was used to quantify the axial resolution of the UST. The measured axial resolution for the 20 MHz transducer ranged from 0.109 mm to 0.137 mm across different distances, while the 10 MHz transducer exhibited a resolution between 0.143 mm and 0.175 mm (**Fig. 6c**). These experimental values were compared to theoretical resolutions, which were calculated as half the spatial pulse length, yielding 0.07 mm for the 10 MHz and 0.0385 mm for the 20 MHz transducers. Discrepancies between the measured and theoretical values are most likely due to the wire diameter; achieving theoretical precision would require a wire diameter of less than 10 μm, which presents significant fabrication challenges.

The final target was a phantom created using a mixture of agar and graphite to mimic tissue, with an embedded cavity simulating a cyst. **Fig. 6d** shows the B-mode image produced by the LIG/PVDF-TrFE UST through the attenuating medium of the agar-graphite mixture. The image clearly shows the distance at which the phantom was positioned within the tank. The first reflection corresponds to the interface between the water and the agar-graphite mixture, and the dark circle with an 8 mm diameter simulates a cyst. The image is comparable to one obtained by the clinical system (**Fig. S14**). Contras-to-noise-ratio (CNR) of the region was computed to be 0.9258 by selecting a region inside and outside the cyst. The image is obtained after high receive gain and multiple pulse averages as described in the Materials and methods. Enhancing the imaging capabilities is possible by applying improved filtering, electrical impedance matching networks to minimize reflections, and time gain compensation (TGC) in the receive chain to mitigate some of the attenuation effects.

Lastly, we demonstrated that the geometry of active US elements is defined by the graphene pattern, which can be precisely controlled by the direct photothermal laser approach. Hydrophone scans of the M-shaped and Doppler devices reveal that the acoustic fields are well-matched to the geometry of the underlying graphene pattern (**Fig. 6e** and **f**), which illustrates the suitability of our fabrication technique for producing wearable transducer arrays with high spatial accuracy and tunable geometry. This technique offers significant potential for advancing the design of customized ultrasound devices for medical and non-medical applications.

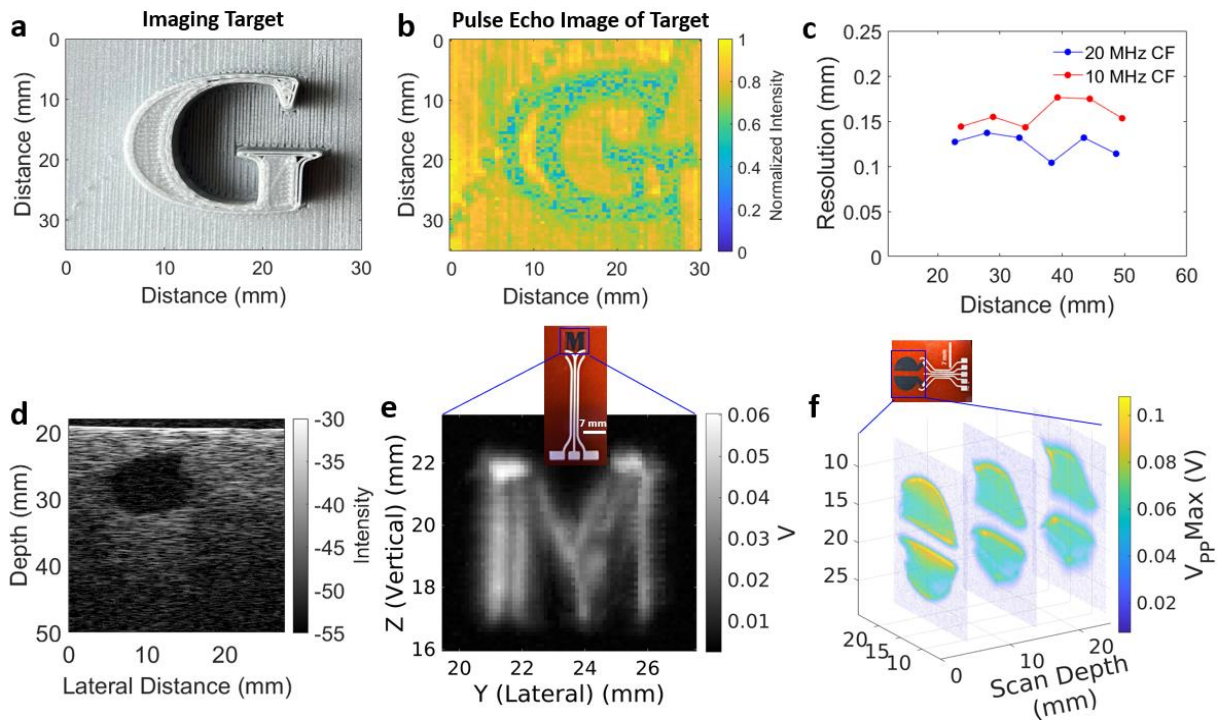

**FIGURE 6.** (a) Photo of the 3D printed "G" target (b) Pulse echo image of the "G" target obtained by LIG/PVDF-TrFE ultrasound transducer. (c) Axial resolution obtained from wire phantom for 20 and 10 MHz LIG/PVDF-TrFE ultrasound transducers. (d) B-mode Image of the phantom in attenuating Agar and graphite mixture produced by the graphene-based transducer. (e) Hydrophone scan of the M-shaped and (f) dual-element transducers.

**Conclusion**

In summary, we proposed an innovative and cost-effective approach for the development of flexible ultrasound transducers by integrating LIG with PVDF-TrFE. Our method leverages laser graphitization to produce LIG electrodes from commercial PI substrate, which are then combined with PVDF-TrFE via 3D printing to create a flexible LIG/PVDF-TrFE ultrasound patch. This technique enables the precise tuning of the central frequency by adjusting the PVDF-TrFE thickness, providing versatility in designing transducers for specific applications. These adjustments allow the transducer to be optimized for different imaging depths and resolutions, making it adaptable to a wide range of clinical and diagnostic needs. The dielectric, piezoelectric properties, and acoustic performance of the LIG/PVDF-TrFE transducer were characterized. We have determined key factors that greatly affect the ultrasound transducer's performance, such as the poling processing of the PVDF-TrFE layer, along with the thicknesses of both the PVDF-TrFE layer and the PI substrate. By fine-tuning the thickness of PVDF-TrFE, we can precisely adjust the central frequency of the transducers, ranging from 10 to 28 MHz. The LIG/PVDF-TrFE ultrasound patch exhibits a high signal amplitude of 6.72 V and SNR of 433, along with a −6 dB bandwidth of 8.86 MHz (37%). The peak-to-peak output voltage of the LIG-based transducer is more than three times greater than that of the silver-based transducer. This improvement is attributed to the higher piezoelectricity of the LIG/PVDF-TrFE composite, which results from the PVDF-TrFE being embedded within the porous structure of the LIG, thereby enhancing the effective surface area of the piezopolymer. The porous structure of LIG increases the contact area between the piezopolymer and the electrode, facilitating more efficient charge transfer and improved polarization under mechanical stress, which significantly enhances the piezoelectric response. Our ultrasound patch produces high-quality two-dimensional images, including B-mode images of cysts. The scalability and patterning of LIG-based electrodes result in the desired sensor configuration with uniform PVDF-TrFE deposition across the device, showcasing the potential of our technique for developing array transducers. This technology holds promising potential beyond traditional medical imaging, with applications in wearable health monitoring, enabling real-time diagnostics for conditions such as cardiovascular health. Furthermore, the advancement offers compact, cost-effective systems that could expand diagnostic imaging capabilities, ultimately improving patient care and encouraging further innovation in both medical and non-medical fields.

**Materials and methods**

*Fabrication of laser-induced graphene from PI.*

A 10.61 μm $CO_2$ laser-cutter system (Universal Laser Systems, VLS2.30) was used for fabricating LIG from commercial polyimide at a laser power of 4.2 W. The pulse duration and scan speed were fixed at 14 μs, and 3.5 in $s^{-1}$, respectively, and the laser treatment was performed under ambient conditions.

*Fabrication of LIG/PVDF-TrFE ultrasound transducers*

Initially, LIG was synthesized as a functional working electrode from commercial polyimide film (125 μm-thick) using photothermal laser treatment. Subsequently, conductive silver ink (FS0142, ACI Materials) was used to print interconnects and contact pads using a printer (Voltera, NOVA). This was followed by the application of UV-cure silicone (SS-5083 UV Dual Cure Acetoxy Silicone, Silicone Solutions) as a shadow mask and deposition of 15 mg PVDF-TrFE (Piezotech FC20 ink, Arkema) to obtain ~50 μm-thick piezoelectric material. Then, the PVDF-TrFE was degassed under a vacuum for three hours. The printed PVDF-TrFE layer underwent drying and annealing at 140°C for 20 minutes. After the annealing process, the PVDF-TrFE-based transducer was positioned on a hot plate for corona-discharge poling. It was subjected to a high DC voltage with a 170 V/μm electric field for 60 minutes at a temperature of 80 °C. The electric field was maintained while the temperature was gradually lowered to 40 °C over 100 minutes. Next, a gold (Au) layer was coated (~60 nm thick) using an ion sputter coater (SPT-20, Coxem) at a current of 3 mA for 300 seconds. To complete the process, a 5 μm thick parylene (Parylene-C dimer) film was deposited on the device for passivation chemical vapor deposition, using a commercial parylene coating system (Labcoter 2, SCS).

*Dielectric and Piezoelectric Characterization*

The electrical impedance of the transducers was measured at room temperature using an impedance analyzer (PSM3750 2 Channel Frequency Response Analyzer, Newtons4th) to obtain resonance and antiresonance frequencies and quantify the electromechanical coupling coefficient. Capacitance and permittivity of the transducers were measured using an LCR meter (IM3536, HIOKI). The piezoelectric coefficients of the transducers were measured using a $d_{33}$ meter (Piezo d33 Test System, ACP International)

*Acoustic Output Measurement*

A pulser/receiver (5073PR, OLYMPUS) was used to obtain the signal amplitude, central frequency, and bandwidth. An aluminum was positioned in the water tank for reflection. The relative position between the transducer and the aluminum was adjusted to obtain the highest signal amplitude. The two-way echo response was directly captured by the same transducer and visualized on an oscilloscope (T3DSO2204A, Teledyne LeCroy). The frequency domain was then analyzed via fast Fourier transform (FFT). The center frequency (CF) and −6 dB bandwidth (BW%) were determined via the following equations:

$$CF = \frac{f_1 + f_2}{2} \qquad 1$$

$$BW(\%) = \frac{f_2 - f_1}{CF} \times 100\% \qquad\qquad 2$$

Where $f_1$ and $f_2$ represent the frequency where the FFT magnitude of the echo drops by −6 dB, with $f_1$ being lower than $f_2$.

All measurements were performed using the following pulser-receiver settings: transmit energy level 4, damping level 1, pulse repetition frequency (PRF) of 200 Hz, and receiver gain of 30 dB.

*Transmit/receive efficiency measurement*

The device was driven with a 20 Vpp sine wave in a 0.5 µs rectangular window using an Agilent 33600A function generator, with the frequency incremented from 10 to 30 MHz in steps of 250 kHz. Acoustic pressure was measured using an Onda HGL-0200 hydrophone, submerged in a water bath and positioned behind an LDPE membrane, with the device placed against the membrane. The driving signal was measured with a high-impedance voltage probe and recorded alongside the acoustic signal from the hydrophone. A 50 mm hydrophone range was used to ensure far-field conditions for frequencies below 25 MHz, taking into account the focusing properties of a 3.5 mm circular element and an expected resonant frequency of approximately 22 MHz from a 50 µm PVDF film. At each frequency, the mean amplitude of the driving voltage and acoustic pressure were calculated within a time window 2.5-4.5 µs from the initial transient. The magnitude of the transmit transfer function ($T_t(\omega)$) for the observed depth was then determined as $T_t(\omega) = P(\omega) V(\omega)^{-1}$, with $P(\omega)$ being the recorded pressure and $V(\omega)$ is the voltage over the transducer electrodes. The receive transfer function ($T_r(\omega)$) was then derived from $T_t$ according to the relationship $T_t(\omega) T_r(\omega)^{-1} = 2 Z(\omega) A \rho_0^{-1} c_0^{-1}$, where $Z(\omega)$ is the device's electrical impedance, A is the area of the active element, and $\rho_0$ and $c_0$ are the density and sound speed of the propagation medium [40].

*Fabrication and imaging of phantoms*

The LIG/PVDF-TrFE UST was driven by a pulser/receiver (5073PR, OLYMPUS) set to a receive gain of 30 dB. The device was positioned in front of a water tank, consistent with previous experiments. Each phantom was securely placed inside the tank and connected to the 3D motorized stage using 3D-printed parts and screws. The first phantom, designed in SolidWorks, consisted of a base plate with the letter "G" extruded outwards, such that the top layer of the "G" extended 5 mm above the base plate. Using a 3D motor stage, the reflector was translated across a 2D plane in 0.1 mm increments. At each step, the radiofrequency data collected from an oscilloscope was processed using an envelope detection technique, which was selected for its ability to improve image clarity by extracting the amplitude of the signal, reducing noise, and enhancing contrast.

The second phantom, also designed in SolidWorks, featured the same base plate with two extruded semi-circular structures, each containing a central hole. An 80 μm diameter wire was threaded through these holes and fixed in place with quick-setting super glue. The phantom was placed in a water tank with an acoustically transparent window and a motorized stage was controlled to change the distance of the phantom from the UST. Initially, the wire was positioned at approximately 2 cm from the UST to optimize the reflected signal. After lateral alignment, the stage moved the wire away from the transducer in 5 mm steps. At each step, the reflected signal was captured with an oscilloscope and resolution was calculated as the full width at half-maximum (FWHM) of the reflected ultrasound pulse from the wire. Two transducers were tested, one with a center frequency of 10 MHz and the other at 20 MHz.

The final phantom was a 7 cm × 7 cm × 7 cm plastic cube, into which a 1 cm diameter rod was inserted approximately 0.4 cm from the surface, running through the entire cube. A solution was prepared by mixing 400 mL of water with 8 gr of graphite and 9 gr of agar following the mixture described in the literature [41]. The mixture was heated on a hot plate while stirring at 350 rpm, then poured into the plastic cube and allowed to set at room temperature. The cube was then filled with a tissue-mimicking solution. After the mixture solidified, the rod was removed to create a cavity. The cavity mimics a cyst or vessel depending on the orientation of the scan. For these scans, a single element UST with a center frequency of 10 MHz was chosen to reduce the effect of attenuation on the signal. A motorized stage was utilized to perform a 2D scan by repositioning the phantom such that the UST scans a different region of the phantom. In order to obtain an image, the gain of the pulsar receiver was set to 60 dB and an averaging of 64 scans was done at each imaging location.

**Data availability**

The data that support the plots within this paper and other findings of this study are available from the corresponding author upon reasonable request.

**CRediT authorship contribution statement**

**Shirin Movaghgharnezhad:** Conceptualization, Methodology, Validation, Investigation, Formal analysis, Writing – original draft, Writing – review & editing. **Ehsan Ansari**: Validation, Investigation, Formal analysis. **Clayton A Baker:** Validation, Investigation,

Formal analysis, Writing – review & editing. **Ahmed A Bashatah:** Investigation, Formal analysis. **Dulcce A Valenzuela:** Investigation. **Pilgyu Kang:** Investigation, Formal analysis, Validation, Supervision, Writing – review & editing. **Parag V Chitnis**: Conceptualization, Methodology, Validation, Investigation, Formal analysis, Resources, Supervision, Project administration, Funding acquisition, Methodology, Writing – review & editing.

**Declaration of Competing Interest**

The authors declare that they have no known competing financial interests or personal relationships that could have appeared to influence the work reported in this paper.

**Acknowledgments**

Effort sponsored by the Government under Other Transaction Number W81XWH-15-9-0001. S.M. acknowledge the Newtons4th company for assisting us with the impedance measurements of the transducer using their impedance analyzer instrument.

**Disclaimer**

The views and conclusions contained herein are those of the authors and should not be interpreted as necessarily representing the official policies or endorsements, either expressed or implied, of the U.S. Government.

**Appendix A. Supplementary data**

Supplementary data to this article can be found online at